\documentclass[reprint, superscriptaddress, secnumarabic, amssymb, nobibnotes, aps, prl]{revtex4-1}

\usepackage{graphicx}
\usepackage{epstopdf}
\usepackage[T1]{fontenc}
\usepackage{amsbsy}
\usepackage{gensymb}
\usepackage[T1]{fontenc}
\usepackage{amsmath}
\usepackage{amssymb}
\usepackage{bbm}
\usepackage{braket}
\usepackage{xcolor}
\allowdisplaybreaks
\usepackage{graphicx}
\usepackage{orcidlink}
\usepackage{hyperref}  
\hypersetup{
    bookmarks=true,         % show bookmarks bar?
    unicode=false,          % non-Latin characters 
    pdftoolbar=true,        % show Acrobat
    pdfmenubar=true,        % show Acrobat 
    pdffitwindow=false,     % window fit to page when opened
    pdfstartview={FitH},    % fits the width of the page to the window
    pdftitle={Stabilization of Ambient Pressure Rocksalt Crystal Structure and High Critical Field Superconductivity in ReC via Mo and W Substitution)},% title
    pdfauthor={},      %author
    pdfsubject={},   % subject of the document
    pdfcreator={},   % creator of the document
    pdfproducer={}, % producer of the document
    pdfkeywords={} {} {}, % list of keywords
    pdfnewwindow=true,      % links in new window
    colorlinks=true,       % false: boxed links; true: colored links
    linkcolor=blue, %red,          % color of internal links (change box color with linkbordercolor)
    citecolor=blue,        % color of links to bibliography
    filecolor=magenta,      % color of file links
    urlcolor=blue           % color of external links
} 

\usepackage[normalem]{ulem}

\newcommand{\figref}[1]{Fig.~\ref{#1}}

\newcommand{\tableref}[1]{Table~\ref{#1}}

\renewcommand{\approx}{\simeq}

\begin{document}
\title{Stabilization of Ambient Pressure Rocksalt Crystal Structure and High Critical Field Superconductivity in ReC via Mo and W Substitution}

\author{{P. K. Meena}\,\orcidlink{0000-0002-4513-3072}}
\affiliation{Department of Physics, Indian Institute of Science Education and Research Bhopal, Bhopal, 462066, India}
\author{S. Jangid}
\affiliation{Department of Physics, Indian Institute of Science Education and Research Bhopal, Bhopal, 462066, India}
\author{R. K. Kushwaha}
\affiliation{Department of Physics, Indian Institute of Science Education and Research Bhopal, Bhopal, 462066, India}
\author{P. Manna}
\affiliation{Department of Physics, Indian Institute of Science Education and Research Bhopal, Bhopal, 462066, India}
\author{S. Sharma}
\affiliation{Department of Physics, Indian Institute of Science Education and Research Bhopal, Bhopal, 462066, India}
\author{P. Mishra}
\affiliation{Department of Physics, Indian Institute of Science Education and Research Bhopal, Bhopal, 462066, India}
\author{{R.~P.~Singh}\,\orcidlink{0000-0003-2548-231X}}
\email[]{rpsingh@iiserb.ac.in}
\affiliation{Department of Physics, Indian Institute of Science Education and Research Bhopal, Bhopal, 462066, India}

\begin{abstract}
Transition-metal-based carbides (TMCs), renowned for their exceptional hardness, mechanical strength, and thermal properties, have recently emerged as promising candidates for topological superconductivity. In this study, we synthesized ReC in the NaCl structure at ambient pressure by substituting Mo or W at the Re-site. We investigated the superconducting properties of Re$_{1-x}$T$_{x}$C (where T = Mo, W) for $x = 0.5$ using magnetization, resistivity and specific heat measurements. These compounds display type-II, fully gapped, weakly coupled superconductivity with high critical fields, establishing them as new members of superconducting ultra-hard materials at ambient pressure and paving the way for superconducting device applications under extreme conditions.

\end{abstract}
\maketitle

%\section{INTRODUCTION}
%\textbf{INTRODUCTION:} 
Transition-metal carbides (TMCs) are a class of refractory, ultra-hard materials with a unique combination of complex covalent, ionic, and metallic bonding, resulting in exceptional mechanical and electronic properties and making TMCs promising candidates for advanced material applications. Similar to 2D materials such as graphene and transition-metal dichalcogenides \cite{TMCs1, TMCs2, Carbides1, Carbides2}. Renowned for their superconductivity, high melting points, outstanding strength, corrosion resistance, and excellent thermal and electrical conductivity, TMCs have found applications in cutting tools, wear-resistant coatings, and catalysts. Their ultra-hardness provides a viable alternative to diamonds, which are often costly and difficult to synthesize \cite{2D-TMCs}. Moreover, the properties of TMCs can be tailored by controlling their valence electron concentration (VEC)\cite{Legar}. However, the synthesis of TMCs can be challenging due to the off-stoichiometric carbon content.

The superconductivity of these refractory materials makes TMCs suitable candidates for device applications in extreme conditions. Recently, rocksalt-structured superconducting TMCs have emerged as prime candidates for hosting topological superconductivity due to their non-trivial band topology \cite{TSC, TopologicalSC, Majorana-fermion}. Some of these carbides, such as XC (where X = Nb, Mo, Ta, V, and Cr) \cite{Nb/TaC, V/CrC, MoC1, TaC-hexagonal1, TaC-2, Dirac-semimetal-NbC, MoC}, exhibit intriguing non-trivial features, including Fermi surface nesting, nodal line, and the semimetallic nature of Dirac \cite{Nodalsemimetal, Fermisurface-nesting}, along with a high superconducting transition temperature (T$_C$) compared to other reported topological superconductors. Isostructural ScS shows unconventional properties with time-reversal symmetry breaking (TRSB) \cite{ScS}. Rocksalt-structured carbides like (Nb, Ta, Mo, W)C show a transition temperature (T$_C$) of up to 14 K \cite{Mo/WC}. Recent theoretical calculations also suggest the possibility of unconventional superconductivity in TMCs \cite{3d-carbides}.

Re-based carbides are particularly notable for their exceptional high-temperature resistance and superior mechanical properties, which surpass those of diamond under high-pressure conditions, making them highly desirable for industrial applications. Although the superconductivity and ultra-incompressibility of ReC have been studied theoretically, stabilizing the rocksalt structure of ReC under ambient conditions remains a challenge \cite{ReC4, ReC, Re4C, ReC1, ReC2, ReC-WC}.

In this paper, we report the stabilization of the NaCl-type structure at ambient pressure and the induction of superconductivity in non-superconducting ReC through the Mo and W substitution. We present a comprehensive study of the structural and superconducting properties of the compounds Re$_{1-x}$T$_{x}$C (where T = Mo and W) for $x = 0.5$. These compounds have been predicted to exhibit remarkable mechanical properties, including ultra-compressibility and superhardness, making them promising candidates for advanced applications \cite{ReWC2-1, ReWC2-2, Mo/WReC2}. The superconductivity of these compounds was examined by resistivity, magnetization, and specific heat measurements, revealing bulk type-II superconductivity with weakly coupled BCS characteristics.

\begin{figure*}
\includegraphics[width=2.05\columnwidth]{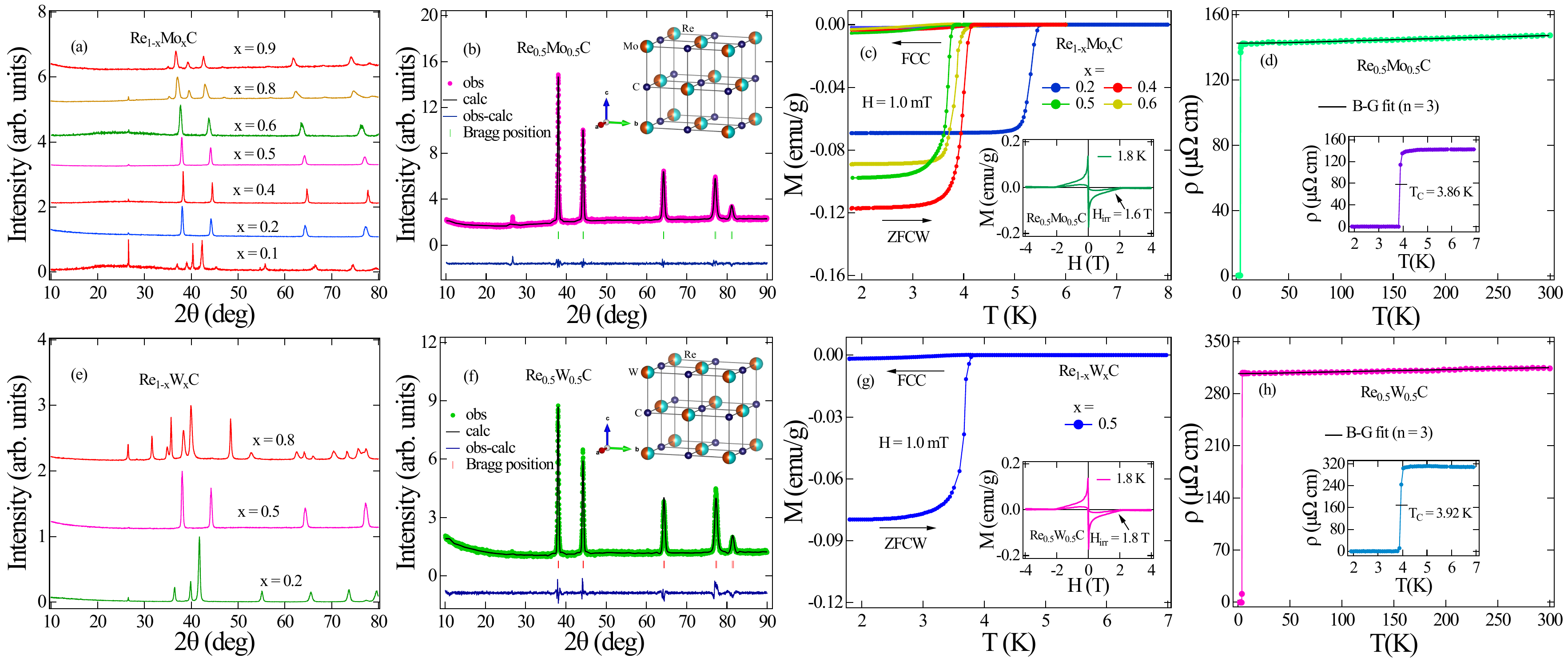}
\caption {\label{Fig1} (a) and (e) Powder XRD patterns for Re$_{1-x}$Mo$_{x}$C ($x$ = 0.1–0.9) and Re$_{1-x}$W$_{x}$C ($x$ = 0.2, 0.5, 0.8). (b) and (f) Rietveld-refined XRD patterns for Re$_{0.5}$Mo$_{0.5}$C and Re$_{0.5}$W$_{0.5}$C, insets shows NaCl-type unit cells. (c) and (g) Temperature-dependent DC magnetization in ZFCW and FCC modes at 1.0 mT shows superconducting transitions; insets show magnetization versus field at 1.8 K. (d) and (h) Temperature-dependent resistivity for Re$_{0.5}$Mo$_{0.5}$C and Re$_{0.5}$W$_{0.5}$C, with insets highlighting T$_{C}$ of 3.86 K and 3.92 K, respectively.}
\end{figure*}

Polycrystalline Re$_{1-x}$T$_x$C (T = Mo, W) was synthesized using an arc melter with high-purity ($4N$) Mo, W, Re, and C in stoichiometric ratios of $1-x:x:1$. The materials were melted in an Ar atmosphere on a water-cooled copper hearth and remelted several times for homogeneity. The crystal structure and phase purity were analyzed by powder X-ray diffraction (XRD) with CuK${\alpha}$ radiation. Magnetic properties were measured with a Quantum Design MPMS-3, while electrical resistivity and specific heat were measured using the four-probe method and the two-tau time relaxation method on a Quantum Design PPMS 9T system.

\begin{table}[b]
\caption{Structural parameters and superconducting T$_{C}$ of Re$_{1-x}$T$_{x}$C (T = Mo, W).}
\label{tbl: lattice parameters}
\setlength{\tabcolsep}{1pt}
%\vspace{-0.05cm}
Re$_{1-x}$Mo$_{x}$C
\vspace{-0.15cm}
\begin{center}
\begin{tabular}[b]{l c c c}
\hline 
\hline
Compound & Phase & Lattice parameters & T$_{C}$ (K) \\
 &  & (a = b = c ($\text{\AA}$)) & \\
\hline
%\\[0.1ex]
ReC \cite{ReC-latticeparameters}& High pressure  & 4.005 &-\\
Re$_{0.9}$Mo$_{0.1}$C& Mixed & -&- \\
Re$_{0.8}$Mo$_{0.2}$C& Single & 4.0115 & 5.51\\
Re$_{0.6}$Mo$_{0.4}$C&  Single &4.0716 &4.15\\
Re$_{0.5}$Mo$_{0.5}$C&  Single &4.1007 & 3.77\\
Re$_{0.4}$Mo$_{0.6}$C&  Single &4.1340 & 4.19\\
Re$_{0.2}$Mo$_{0.8}$C&  Mixed & - & - \\
Re$_{0.1}$Mo$_{0.9}$C&  Mixed & -& - \\
MoC \cite{Mo/WC}& Single  & 4.270 &14\\
%\\[0.1ex]
\hline
\end{tabular}
\par\medskip\footnotesize
\end{center}
\vspace{-0.5cm}
Re$_{1-x}$W$_{x}$C
\vspace{-0.15cm}
\begin{center}
\begin{tabular}[b]{l c c c}
\hline 
\hline
Compound & Phase & Lattice parameters & T$_{C}$ (K) \\
 &  & (a = b = c ($\text{\AA}$)) & \\
\hline
%\\[0.1ex]
ReC \cite{ReC-latticeparameters} & High pressure  & 4.005 &-\\
Re$_{0.8}$W$_{0.2}$C & Mixed  & - & -\\
Re$_{0.5}$W$_{0.5}$C& Single &4.0933 & 3.78\\
Re$_{0.2}$W$_{0.8}$C& Mixed & - & - \\
WC$_{1-x}$ \cite{Mo/WC, WC1-x} & Single  & 4.2387 &-\\
%\\[0.1ex]
\hline
\end{tabular}
\end{center}
\end{table}

The XRD patterns of Re$_{1-x}$T$_{x}$C compounds (where T = Mo, W) with Mo: $x$ = 0.1, 0.2, 0.4, 0.5, 0.6, 0.8, 0.9) and W: $x$ = 0.2, 0.5, 0.8) are presented in \figref{Fig1}(a) and (e). It suggests a single-phase NaCl structure stabilized between $x$ = 0.2 - 0.6 for Mo and $x$ = 0.5 for the substitution of W. The structural refinement of the XRD patterns was performed using a Rietveld method using FULLPROF software \cite{FULLPROF}, a NaCl-type face-centered cubic crystal structure with space group $Fm\bar{3}m$ (no. 225) and the phase purity of the sample. The refined XRD patterns for Re$_{0.5}$T$_{0.5}$C (T = Mo, W) are shown in \figref{Fig1}(b) and (f). The inset illustrates the NaCl-type structure, with mixed T and Re atoms at the 4a (0, 0, 0) site and carbon atoms at the 4b (0.5, 0.5, 0.5) site.
The lattice parameters, listed in Table~\ref{tbl: lattice parameters}, are consistent with published data for compositions $x$ = 0.5 \cite{Mo/WReC2} and slightly smaller than those of binary carbides \cite{Mo/WC}. This reduction is attributed to the smaller atomic radius of Re. The observed instability with 20\% W substitution compared to Mo substitution might be due to the inherently stable NaCl structure of MoC and the intrinsically unstable NaCl structure of WC under normal pressure. These findings emphasize the complex relationship between substitution levels and structural stability in these materials.

\figref{Fig1}(c) and (g) show magnetization measurements in zero-field-cooled warming (ZFCW) and field-cooled cooling (FCC) modes, with a 1.0 mT applied field, confirmed superconductivity in both compounds. The superconducting transition temperatures (T$_{C}$) are listed in Table~\ref{tbl: lattice parameters}, for Re$_{1-x}$T$_{x}$C (T = Mo, W). The stronger diamagnetic response in FCC compared to ZFCW due to flux trapping confirms the type-II nature of these superconductors. The superconducting fractions exceed 100\% due to the irregular sample shape and demagnetization effects \cite{demagnetisation-factor}. The hysteresis magnetization loops for $x$ = 0.5 compositions (insets in \figref{Fig1}(c) and (g)) further support type-II superconductivity, with H$_{irr}$ values of 1.6 and 1.8 T for Mo and W-based carbides, respectively.

Figs.~\ref{Fig1}(d) and (h) show the temperature-dependent AC electrical resistivity $\rho(T)$ of Re$_{0.5}$T$_{0.5}$C (T = Mo, W) carbides, measured from 1.9 to 300 K in a zero magnetic field. Above the superconducting transition temperature (T$_{C}$), $\rho(T)$ remains nearly constant, indicating poor metallic behavior. The residual resistivity ratio (RRR = $\rho_{300}$/$\rho_{10}$) for the Mo and W variants, 1.03(7) and 1.01(8), respectively, indicating high intrinsic disorder in the compounds. These values are similar to those reported for binary carbides and high-entropy alloys. The inset of Figs.~\ref{Fig1}(d) and (h) shows a drop in resistivity at T$_{C}$ = 3.86(4) K for Re$_{0.5}$Mo$_{0.5}$C and 3.92(7) K for Re$_{0.5}$W$_{0.5}$C, confirming superconductivity. The observed T$_{C}$ values are consistent with the magnetization measurement.\\
%In various cubic compounds (A15, B1, A12, and D5c, the relationship between T$_{C}$ and valence electron count (VEC) found. Our compounds crystallize in the rocksalt (B1) structure, common for many nitrides and carbides. The shared electron per atom (e/a) and lattice constant between the two compounds might be the primary factors influencing their superconducting properties\cite{Legar}. T$_{C}$ generally depends on e/a, increasing from below 0.01 K at e/a = 4 (TiC, ZrC, HfC) to above 14 K at e/a = 5. Our compounds, with e/a = 5.25, exhibit T$_{C}$ values around 4 K \cite{Mo/WReC2}. 
The resistivity data above T$_{C}$ can be well understood by the Bloch-Gruneisen (BG) model with n = 3 expressed as \cite{BG-Model},
\begin{equation}
{\rho(T)}= \rho_{0} + A \left(\frac{T}{\theta_{D}}\right)^3 \int_{0}^{{\theta_{D}}/T} \frac{x^{3}}{(e^{x}-1)(1-e^{-x})} dx
\label{eqn1:BG}
\end{equation}
where $\rho_{0}$ is the residual resistivity, A is the degree of electron correlation in the substance \cite{n} and $\theta_{D}$ is the Debye temperature. The fitted resistivity data from the BG model provides A = 2.27 $\mu \Omega.cm$, $\theta_{D}$ = 283.06, $\rho_{0}$ = 142.64 $\mu \Omega.cm$ for Re$_{0.5}$Mo$_{0.5}$C and A = 2.32 $\mu \Omega.cm$, $\theta_{D}$ = 287.06, $\rho_{0}$ = 307.12 $\mu \Omega.cm$ for Re$_{0.5}$W$_{0.5}$C. These values of $\theta_{D}$ are relatively close to those obtained from the heat capacity data (discussed later).

\begin{figure}
\includegraphics[width=0.98\columnwidth]{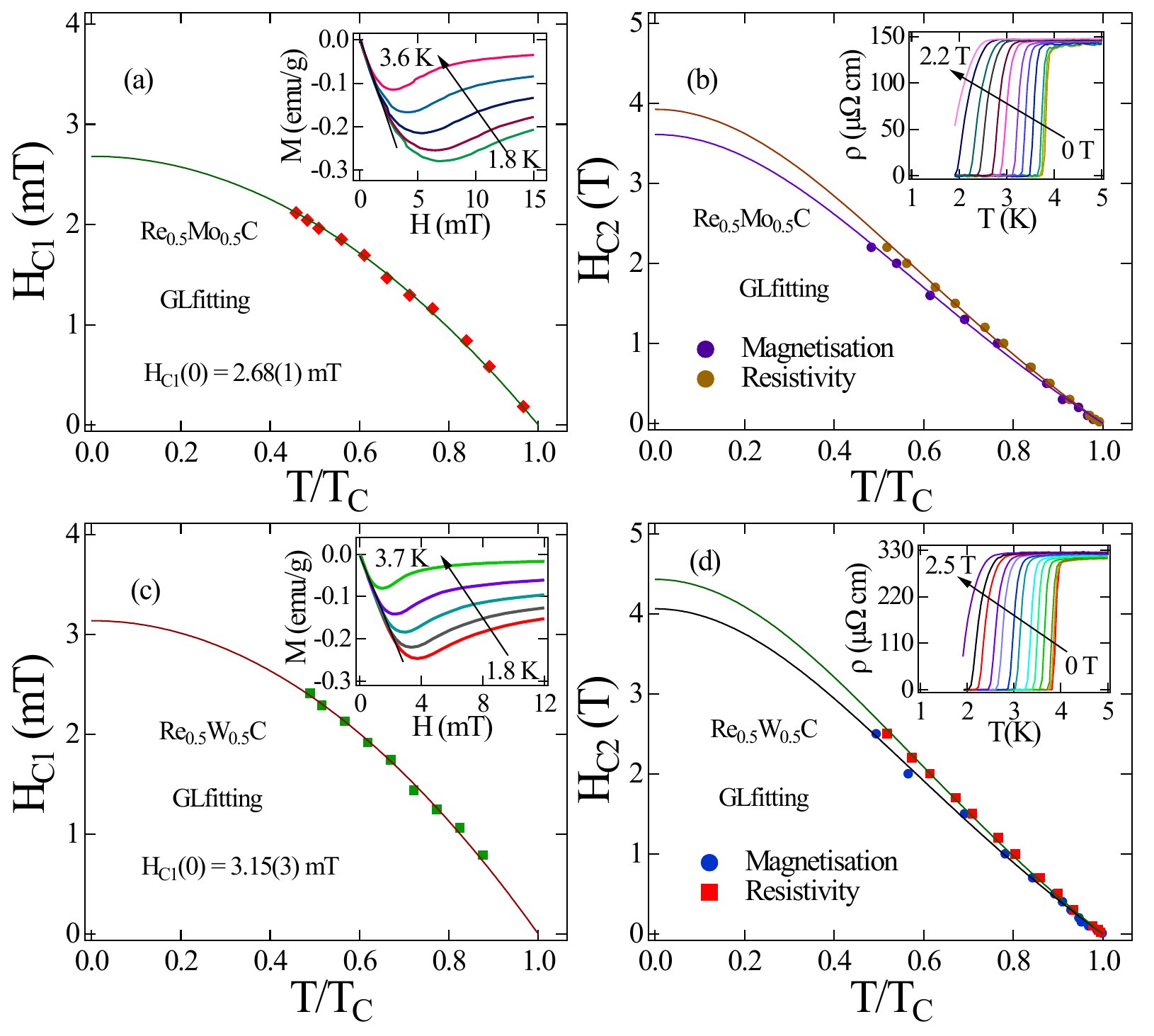}
\caption {\label{Fig3} (a) and (c) Temperature dependence of the lower critical field fitted using the GL model, with insets showing magnetization variation under low magnetic fields. (b) and (d) Upper critical fields as a function of temperature, fitted with the GL equation, with insets displaying resistivity data under varying magnetic fields.}
\end{figure}

Low-field magnetization curves were recorded at various temperatures to determine the lower critical field, H$_{C1}$(0). The onset of vortex formation and the mixed state was identified as the deviation point from the Meissner line in the M-H curves (insets in Figs.~\ref{Fig3}(a) and (c)). The temperature-dependent H$_{C1}$ data was analyzed using the Ginzburg-Landau (GL) equation, expressed as: $H_{C1}(T) = H_{C1}(0)\left[1-\left(\frac{T}{T_{C}}\right)^{2}\right]$ providing H$_{C1}$(0) values of 2.68(1) and 3.15(3) mT for Re$_{0.5}$T$_{0.5}$C (T = Mo, W), respectively.

To calculate the upper critical field, H$_{C2}$(0), the effect of the applied magnetic field on T$_{C}$ was measured through magnetization and electrical resistivity (insets in Figs.~\ref{Fig3}(b) and (d)). The temperature-dependent H$_{C2}$ data was analyzed using the GL equation, expressed as: $H_{C2}(T) = H_{C2}(0)\left[\frac{1-t^{2}}{1+t^{2}}\right]$, yielding H$_{C2}$(0) values of 3.60(3) and 3.92(2) T for Re$_{0.5}$Mo$_{0.5}$C, and 4.06(4) and 4.42(3) T for Re$_{0.5}$W$_{0.5}$C, from magnetization and resistivity measurements, respectively.

Temperature-dependent H$_{C1}$(0) and H$_{C2}$(0) plots, along with GL fits, are shown for both samples in Figs.~\ref{Fig3}(a) and (c) and Figs.~\ref{Fig3}(b) and (d), respectively. The obtained H$_{C2}$(0) and H$_{C1}$(0) values were used to calculate the superconducting characteristic length scales, coherence length $\lambda_{GL}$ and penetration length $\xi_{GL}$ \cite{Tin1, Tin2}. The expressions for these parameters are given by:
\begin{equation}
H_{C2}(0) = {\frac{\Phi_{0}}{2\pi \xi_{GL}^2}}.
\label{eqn6:lamda}
\end{equation}
\begin{equation}
H_{C1}(0) = \frac{\Phi_{0}}{4\pi\lambda_{GL}^2}\left( ln \frac{\lambda_{GL}}{\xi_{GL}} + 0.12\right).
\label{eqn6:lamda}
\end{equation}
where $\Phi_{0}$ is the magnetic flux quantum. 
The obtained parameters $\lambda_{GL}$ and $\xi_{GL}$ for the Mo-variant are 5047(5) \text{\AA} and 95.6(6) \text{\AA}. For the W variant, they are 4592(7) \text{\AA} and 90.1(8) \text{\AA}. The derived GL parameters $\kappa_{GL}$ (= $\lambda_{GL}$/$\xi_{GL}$) are 52.20(3) and 50.98(5) for Re$_{0.5}$Mo$_{0.5}$C and Re$_{0.5}$W$_{0.5}$C. These values exceed 1 /$\sqrt{2}$, confirming these superconductors' type II nature. Furthermore, using H$_{C1}$(0), H$_{C2}$(0) and $\kappa_{GL}$, the thermodynamic critical field H$_{C}$ was calculated using the relation: $H_{C}^2 ln[k_{GL}(0)+0.08] = {H_{C1}(0) H_{C2}(0)}$ \cite{Tin3}, yielding values of 48.8(6) and 57.1(3) mT for the Mo and W-based carbides, respectively.

\begin{figure}
\includegraphics[width=0.98\columnwidth]{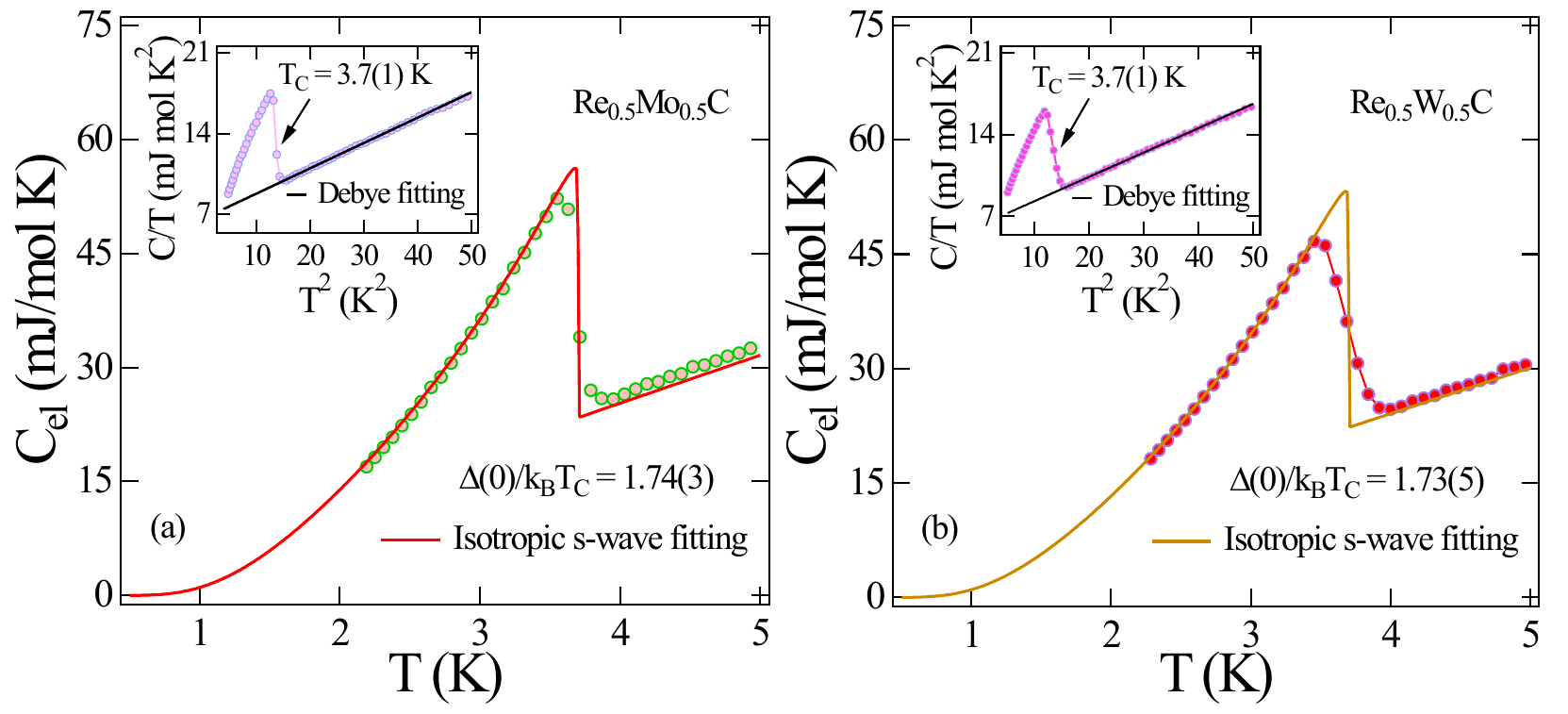}
\caption {\label{Fig4} An isotropic, fully gapped model describes specific heat data for (a) Re$_{0.5}$Mo$_{0.5}$C and (b) Re$_{0.5}$W$_{0.5}$C. The inset confirms superconductivity with a clear anomaly at 3.7(1) K for both carbides.}
\end{figure}

Superconductivity can be destroyed by orbital and Pauli paramagnetic limiting field effects. The orbital limit, H$_{C2}^{orb}$(0), is associated with the kinetic energy of the Cooper pair. The Werthamer-Helfand-Hohenberg (WHH) theory provides the expression \cite{WHHM1, WHHM2}:
\begin{equation}
H^{orb}_{C2}(0) = -\alpha T_{C} \left.{\frac{dH_{C2}(T)}{dT}}\right|_{T=T_{C}}. 
\label{eqn4:WHH}
\end{equation}
where $\alpha$ is the purity factor (0.69 for dirty, 0.73 for clean limits). The initial slope, -$dH_{C2}(T)/{dT}|_{T=T_{C}}$, was obtained from the H$_{C2}$-T phase diagram (Figs.~\ref{Fig4}(b) and (d)). The slope values obtained are 0.96(7) and 0.78(6) T / K for carbides doped with Mo and W, respectively. For dirty limit superconductors ($\alpha$ = 0.69), the calculated H$_{C2}^{orb}$(0) values are 2.54(7) and 2.06 (9) T for carbides based on Mo and W.

The Pauli paramagnetic limiting field, H$_{C2}^{P}$(0), arises from the Zeeman splitting and is given by: H$_{C2}^{P}$(0) = 1.86*T$_{C}$ for weakly coupled BCS superconductors \cite{Pauli1, Pauli2}. The estimated H$_{C2}^{P}$ values are 7.01(2) and 7.03(1) T for Mo and W-based carbides. Since the upper critical field is lower than the Pauli limiting field, the orbital limiting field is responsible for breaking the Cooper pairs in both compounds.

To quantify spin paramagnetic effects, we calculated the Maki parameter, ($\alpha_{m}$), using the expression $\alpha_{m}$ = $\sqrt{2}$ H$_{C2}^{orb}$(0)/H$_{C2}^{P}$(0). The values of $\alpha_{m}$ are 0.51(2) and 0.41(4) for Mo and W variants, respectively.

The Ginzburg number G$_{i}$, a ratio of thermal energy to condensation energy related to the coherence volume, is given by:
\begin{equation}
    G_{i} = \frac{1}{2}\left[\frac{k_{B} \mu_{0} \tau T_{C}}{4 \pi \xi^{3}_{GL}(0) H^{2}_{C}(0)}\right]^{2},
\end{equation}
where $\tau$ is an anisotropic ratio and assumes 1 for cubic structure. Using the values of T$_{C}$, $\xi_{GL}(0)$ and H$_{C}$, we calculated G$_{i}$ = 3.15(1) and 2.43(2) $\times 10^{-6}$ for Mo and W-based carbides. These values are lower than those of high-T$_{C}$ cuprate superconductors (10$^{-2}$) \cite{Gi-1} but higher than those of low-T$_{C}$ superconductors (10$^{-8}$) \cite{Gi-2} suggesting that weak thermal fluctuations contribute to vortex unpinning in these carbides \cite{Gi-3}.

Specific heat measurements were performed in zero magnetic field for Re$_{0.5}$T$_{0.5}$C (T = Mo, W), as shown in Figs.~\ref{Fig4}(a) and (b). A clear transition from the normal to the superconducting state was observed at T$_{C}$ = 3.7(1) K for both Mo- and W-based carbides, consistent with magnetization and resistivity measurements. The Debye-Sommerfeld relation, $C/T$ = $\gamma_{n}$ + $\beta_{3}T^{2}$, where $\gamma_{n}$ is the Sommerfeld coefficient and $\beta_{3}$ is the phononic coefficient $\beta_{3}$, used to analyze normal state behavior. By extrapolating the normal state behavior to low temperatures, we obtained $\gamma_{n}$ and $\beta_{3}$ are 6.60(2) mJ/mol K$^{2}$ and 0.219(8) mJ/mol K$^{4}$ for Re$_{0.5}$Mo$_{0.5}$C and 6.18(3) mJ/mol K$^{2}$ and 0.209(3) mJ/mol K$^{4}$ for Re$_{0.5}$W$_{0.5}$C, respectively. $\beta_{3}$, was used to calculate the Debye temperature, $\theta_{D}$ using the relation $\theta_{D} = \left(\frac{12\pi^{4} R N}{5 \beta_{3}}\right)^{\frac{1}{3}}$, where N is the number of atoms per formula unit and R = 8.314 J mol$^{-1}$ K$^{-1}$ is a gas constant. The calculated values for $\theta_{D}$ are 328(2) and 333(7) K for Mo and W-based carbides, consistent with resistivity data. The Sommerfeld coefficient $\gamma_{n}$, is related to the density of states at the Fermi energy, D$_{C}$(E$_{F}$), as $\gamma_{n}$ = $\left(\frac{\pi^{2} k_{B}^{2}}{3}\right)$ $D_{C}(E_{F})$. Here, $k_{B}$ = 1.38 $\times$ 10$^{-23}$ J K$^{-1}$ is the Boltzmann constant. The calculated values of $D_{C}(E_{F})$ are 2.79(8) and 2.62(2) states eV$^{-1}$ f.u.$^{-1}$ for Re$_{0.5}$T$_{0.5}$C (T = Mo, W respectively). These values are higher than the topological isostructural rock salt carbides (Nb, Ta)C \cite{Nb/TaC, Dirac-semimetal-NbC}. 

McMillan's theory \cite{el-ph} provides a formula to calculate the dimensionless electron-phonon coupling parameter $\lambda_{ep}$, which represents the strength of the attractive interaction between electrons and phonons. The expression is defined as:
\begin{equation}
T_{C} = \frac{\theta_{D}}{1.45} \exp\left[{\frac{1.04(1+\lambda_{ep})}{\mu^{*} (1+0.62.\lambda_{ep})-\lambda_{ep}}}\right];
\label{eqn8:Lambda}
\end{equation}
where $\mu^{*}$ is the repulsive-screened Coulomb potential parameter (0.13 for intermetallic compounds). The estimated $\lambda_{ep}$ values for the Mo and W-variants are 0.574(2) and 0.571(8), respectively, indicating weak-coupled superconductivity.

Specific heat measurements also provide insights into the superconducting gap symmetry. By subtracting the phononic terms from the total specific heat at zero fields, we calculated the electronic contribution $C_{el}$. The specific heat jump $\Delta(0)/\gamma_{n}T_{C}$ was found to be 1.30 (2) and 1.29 (4) for the Mo- and W-based carbide, slightly lower than the BCS value of 1.43. Low-temperature $C_{el}$ was theoretically fitted by the $\alpha$-model of the isotropic BCS weak coupling superconductor \cite{BCS}, represented by the equation:
\begin{equation}
\frac{S}{\gamma_{n} T_{C}}= -\frac{6}{\pi^{2}} \left(\frac{\Delta(0)}{k_{B} T_{C}}\right) \int_{0}^{\infty} \left[ {fln(f)+(1-f)ln(1-f)} \right] dy
\label{BCS}
\end{equation}
where $f(\xi)$ = $(\exp(E(\xi)/k_{B}T)+1)^{-1}$ is the Fermi function, and y = $\xi$/$\Delta(0)$, with $E(\xi$) = $\sqrt{\xi^{2}+\Delta^{2}(t)}$, which is the energy of normal electrons relative to the Fermi energy, where t = T/T$_{C}$ is normalized temperature. The normalized temperature-dependent gap function is written as $\Delta(t)=tanh[1.82((1.018(1/t))-1)^{0.51}]$. Using the relation between entropy $S$ and $C_{el}$, given as C$_{el}$ = $tdS/dt$, we calculated the electronic specific heat. The superconducting gap ($\alpha$ = $\Delta(0)$/$k_{B}T_{C}$), was found to be 1.74(3) and 1.73(5) for carbides based on Mo and W, which is close to the BCS value for isotropic, weakly coupled superconductors, suggesting phonon-mediated superconductivity in these ultra-hard materials. The $\alpha$-model of the isotropic BCS superconductor's best fitting of Figs.~\ref{Fig4}(a) and (b) provides us with this ratio. Data points up to T$_{C}$/10, and microscopic measurements are required to determine the exact pairing mechanism.

To study the electronic properties of these Re-based carbides, we employed a set of equations considering a spherical Fermi surface. These equations allow us to calculate the mean free path, $l_{e}$, and the effective mass, $m^{*}$ \cite{5-equations}. The equations are as follows:
\begin{equation}
\gamma_{n} = \left(\frac{\pi}{3}\right)^{2/3}\frac{k_{B}^{2}m^{*}V_{\mathrm{f.u.}}n^{1/3}}{\hbar^{2}N_{A}}
\label{eqn17:gf}
\end{equation}
\begin{equation}
\textit{l}_{e} = \frac{3\pi^{2}{\hbar}^{3}}{e^{2}\rho_{0}m^{*2}v_{\mathrm{F}}^{2}}, n = \frac{1}{3\pi^{2}}\left(\frac{m^{*}v_{\mathrm{F}}}{\hbar}\right)^{3}
\label{eqn18:le,n}
\end{equation}
where $k_{B}$, $V_{f.u.}$, and $N_{A}$ are the Boltzmann constant, the volume of a formula unit, and the Avogadro number, respectively. For a dirty limit superconductor, the effective magnetic penetration depth $\lambda_{GL}$(0) can be expressed in terms of the London penetration depth $\lambda_{L}$(0) as: 
\begin{equation}
\lambda_{GL}(0) = 
\lambda_{L}
\left(1+\frac{\xi_{0}}{\textit{l}_{e}}\right)^{1/2}, \lambda_{L} =
\left(\frac{m^{*}}{\mu_{0}n e^{2}}\right)^{1/2}
\label{eqn20:f}
\end{equation}
\begin{equation}
\frac{\xi_{GL}(0)}{\xi_{0}} = \frac{\pi}{2\sqrt{3}}\left(1+\frac{\xi_{0}}{\textit{l}_{e}}\right)^{-1/2}
\label{eqn21:xil}
\end{equation}

\begin{figure}
\includegraphics[width=0.96\columnwidth]{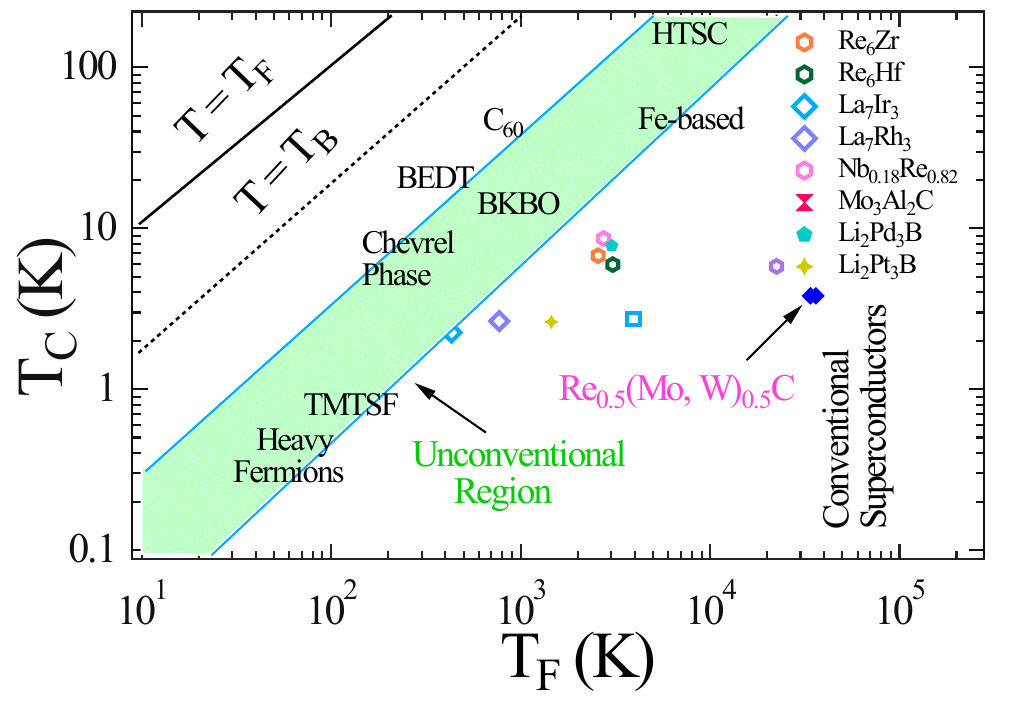}
\caption {\label{Fig5} The Uemura plot for Re$_{0.5}$T$_{0.5}$C (T = Mo, W) is indicated by a blue marker positioned near the conventional superconductivity range.}
\end{figure}

The high value of $\xi_{0}/l_{e}$ suggests that both Re-based carbides have a dirty limit superconductivity. The calculated electronic parameters are summarized in \tableref{tbl: parameters}. 

The Uemura plot helps categorize superconductors as conventional or unconventional based on Fermi temperature and superconducting transition temperature \cite{CS/US}. We calculated the Fermi temperature for these carbides using the equation \cite{Tf}:
\begin{equation}
 k_{B}T_{F} = \frac{\hbar^{2}}{2}(3\pi^{2})^{2/3}\frac{n^{2/3}}{m^{*}}.
\label{eqn13:tf}
\end{equation}

The values obtained for carbides based on Mo and W are T$_{F}$ are 3.38(6) and 3.61(4) $\times 10^{4}$ K. The corresponding T$_{C}$/T$_{F}$ ratios, 0.00011 and 0.00010, fall outside the unconventional band range (0.01 < T$_{C}$/T$_{F}$ < 0.05) in the Uemura plot, indicating that these carbides are conventional superconductors.

\begin{table}
\caption{Superconducting and normal state parameters of Re$_{0.5}$T$_{0.5}$C (T = Mo, W), measured from magnetization, resistivity, and specific heat measurements.}
\label{tbl: parameters}
\setlength{\tabcolsep}{6pt}
\begin{center}
\begin{tabular}[b]{l c c c }\hline \hline
Parameters& unit& Re$_{0.5}$Mo$_{0.5}$C & Re$_{0.5}$W$_{0.5}$C\\
%\\[0.1ex] 
\hline
%\\[0.1ex] 
VEC& -& 5.25& 5.25\\
T$_{C}$ & K &3.77(8) & 3.78(5) \\
H$_{C1}(0)$ & mT & 2.68(1) & 3.15(3) \\ 
H$_{C2}^{mag}$(0) & T & 3.60(3) &4.06(4) \\
H$_{C2}^{res}$(0) & T & 3.92(2) &4.42(3) \\
H$_{C2}^{P}$(0) & T & 7.01(2) &7.03(1) \\
H$_{C2}^{orb}$(0) & T & 2.54(7) &2.06(9) \\
H$_{C}$ & mT  & 48.8(6) & 57.1(3)\\
$\xi_{GL}$& \text{\AA} &95.6(6)&90.1(8) \\
$\lambda_{GL}^{mag}$& \text{\AA} &5047(5)& 4592(7)\\
$G_{i}$ & 10$^{-6}$ & 3.15(1) &2.43(2)\\
$\gamma_{n}$& mJ/mol K$^{2}$&6.60(2) &6.18(3) \\
$\theta_{D}$& K& 328(2) &333(7)\\
$\frac{\Delta(0)}{k_{B}T_{C}}$ (sp) & - & 1.74(3) &1.73(5)\\
$\lambda_{e-ph}$ & - &0.574(2)& 0.571(8)\\
n & $10^{28} m^{-3}$ &1.22(2) & 1.22(2)\\
$m^{*}$ & -&0.66(4) &0.62(2)\\
$v_{F}$ & $10^{5}$ $ms^{-1}$ &12.24(9) & 13.46(4)\\
$\xi_{0}/l_{e}$ & - & 167.4(5)& 147.2(6)\\
T$_{F}$ & $10^{4}$ K &3.38(6) & 3.61(4)\\
%\\[0.1ex]
\hline
\hline
\end{tabular}

\end{center}
\end{table}

%\vspace{-0.4cm}

%The band structure of binary rocksalt carbides (Nb, Ta)C reveals dominant d-electron contributions from transition metals, albeit hybridized, with modest C p-electron involvement. Strong spin-orbit coupling (SOC) in these carbides splits six-fold degenerate points at $\Gamma$ into one doubly degenerate point and one four-fold degenerate point below the Fermi level, contrasting with NbC's less pronounced SOC splitting. The MoC compound exhibits instability with theoretically predicted imaginary phonon bands in its pristine form. However, theoretical and experimental evidence suggests that hole doping eliminates these imaginary bands, stabilizing the lattice \cite{MoC}. This highlights the potential impact of high SOC elements like Re in our systems, suggesting a significant SOC influence on band structure splitting. The lower T$_{C}$ in these carbides compared to binary (Mo, W)C also prompts investigation into the superconductivity dominance of Re, with anticipated SOC effects similar to those of compounds studied theoretically and experimentally (Nb, Ta) C \cite{Mo/WC, Nb/TaC}.

In conclusion, we successfully synthesized ReC, a rocksalt-type crystal structure, at ambient pressure through Mo and W doping. Magnetization, electrical resistivity, and specific heat measurements revealed bulk type-II superconductivity in Re$_{0.5}$T$_{0.5}$C (T = Mo, W) with T$_{C}$ $\approx$ 3.8 K. These compounds exhibit face-centered-cubic NaCl-type structures and weak-coupling s-wave superconductivity. The superconducting properties of Re$_{0.5}$T$_{0.5}$C are comparable to those observed in Mo and W-based borides with the same valence electron count \cite{Mo-WReB}. Further exploration of their mechanical properties is warranted. The combination of potential non-trivial band topology, high spin-orbit coupling elements, enhanced hardness, and high critical field superconductivity in Re$_{0.5}$T$_{0.5}$C provides valuable insights into the relationships between these properties and their potential for practical applications in extreme conditions. Given the time-reversal symmetry breaking observed in ScS \cite{ScS}, these Re-based carbides may offer insights into the frequent occurrence of unconventional superconducting ground states in Re-based superconductors. Additionally, this study paves the way for the development of ultra-hard superconducting materials by exploring the formation of high-entropy carbides through the addition of suitable elements.

%\section{Acknowledgments} 
\textbf{Acknowledgments:} P.K.M. acknowledges the financial support provided by the CSIR, Government of India, through the SRF Fellowship (Award No: 09/1020(0174)/2019-EMR-I). R.P.S. is thankful to the SERB, Government of India, for the Core Research Grant (CRG/2023/000817).

\end{document}